\documentclass [12pt] {article}
\usepackage{graphicx}
\usepackage{epsfig}
\begin {document}
\vskip 1.cm

\begin{center}

{\bf\Large Inelastic shadowing of secondary $\pi^{\pm}$, $K^{\pm}$, $p$, and $\bar{p}$\\
produced in $p+$}{\Large Pb} {\bf\Large collisions at LHC energy\footnote{Presented by C.~Merino
(carlos.merino@usc.es) at the Low-$x$ Meeting 2017, Bisceglie, near Bari (Italy),
June 13$^{th}$-18$^{th}$, 2017.}}\\
\vspace{.5cm}
{\bf G.H. Arakelyan$^a$, C. Merino$^b$, Yu.M. Shabelski$^c$ and A.G. Shuvaev$^c$}\\
\vspace{.5cm}

$^a$A.Alikhanyan National Scientific Laboratory
(Yerevan Physics Institute)\\
Yerevan, 0036, Armenia\\
\vspace{.2cm}

$^b$Departamento de F\'\i sica de Part\'\i culas, Facultade de F\'\i sica \\
and Instituto Galego de F\'\i sica de Altas Enerx\'\i as (IGFAE) \\
Universidade de Santiago de Compostela, Galiza, Spain \\
\vspace{.2cm}

$^{c}$Petersburg Nuclear Physics Institute \\
NCR Kurchatov Institute \\
Gatchina, St.Petersburg 188300 Russia \\
\vskip 0.5 cm

\end{center}

\begin{abstract}
The inclusive spectra of secondaries produced in soft (minimum-bias)
$p$+Pb collisions at LHC energy are calculated in the frame of the
Quark-Gluon String Model, by including the inelastic screening corrections
(percolation effects). These effects are expected to be quite large
at the very high energies, and they should drive down the spectra
in the midrapidity region more than 2 times, at $\sqrt{s_{NN}}$=5~TeV.
\end{abstract}

PACS. 25.75.Dw Particle and resonance production

\section{Introduction}

We compare the results obtained in the frame of the
Quark-Gluon String Model~(QGSM)
with the experimental data for the inclusive
densities of different secondaries obtained by the
CMS collaboration for $p$+Pb at $\sqrt{s_{NN}}$=5~TeV
(see~\cite{paper} for a more detailed version of this work).

In~\cite{CKTr,MPS} it was shown that in the frame of the QGSM
one can obtain a reasonable description of the experimental data
on the inclusive spectra of secondaries produced in d+Au collisions
at $\sqrt{s_{NN}}$=200~GeV (RHIC), by accounting of the
inelastic corrections, which are related to the multipomeron interactions.
These corrections lead to the saturation of
the inclusive density of secondary hadrons in the soft (low $p_T$)
region, where the methods based on perturbative QCD cannot be used.
The effects of the inelastic shadow corrections should increase with
the initial energy, becoming large at the LHC energies.

In principal, two possibilities exist to explain the origin of
the inelastic nuclear screening: either it comes from the diagrams
with Pomeron interactions, or from the interactions
of the produced secondaries with another hadrons.
In the first case, the inelastic screening effects
should be the same for different secondaries, while for the second one
they should be different.

\section{QGSM inclusive spectra of secondary hadrons
with inelastic screening effects in p+A collisions at very\newline
high energies}

In order to produce quantitative results for the inclusive spectra of secondary hadrons,
a model for multiparticle production is needed. It is for that purpose that we have
used the QGSM~\cite{KTM,Kaid} in the numerical calculations presented below.
QGSM is based on the Reggeon calculus and on the $1/N_c$ (or $1/N_f$) expansion in QCD,
where $N_c$ and $N_f$ are the numbers of colors and light flavors, respectively.

Both the high energy hadron-nucleon and hadron-nucleus interactions
are treated in the QGSM as proceeding via the exchange of one
or several Pomerons. The elastic and inelastic processes result
from cutting through or between Pomerons~\cite{AGK}. Each Pomeron
corresponds to a quark-gluon cylinder diagram.
The cut through the cylinder produces two showers of secondaries
(color strings)~\cite{Artru}.
The decay of these strings generates
new quark-antiquark pairs that lead then to the production
of secondary hadrons.

For the nucleon target, the inclusive density $dn/dy$ of a
secondary hadron $h$ has the form~\cite{KTM}:
\begin{equation}
\frac{dn}{dy} =  \frac{1}{\sigma_{inel}}\cdot\frac{d\sigma}{dy}
= \frac{x_E}{\sigma_{inel}}\cdot\frac{d\sigma}{dx_F}
=\sum_{n=1}^{\infty}w_{n}\cdot\phi_{n}^{h}(x)\ \ ,
\end{equation}
where the functions $\phi_{n}^{h}(x)$ determine the contribution
of diagrams with $n$ cut Pomerons, and $w_{n}$ is the probability
for this process to occur~\cite{TM}. Here we neglect the diffractive
dissociation contributions, since it would only be significant in
the fragmentation regions, i.e at large $x_F$.

The specific form of the functions $\phi_{n}^{h}(x)$ is given
by the convolution of the diquark and quark distributions with the
fragmentation functions, both being
determined by Regge asymptotics~\cite{Kai,Sh}.

The probabilities $w_{n}$ in Eq.~(1) are the ratios
of the cross sections corresponding to $n$ cut Pomerons,
$\sigma^{(n)}$, to the total non-diffractive inelastic $pp$
cross section, $\sigma_{nd}$~\cite{TM}.

The contribution of multipomeron exchanges in high energy
$pp$ interactions results in a broad distribution of $w_n$ (see~\cite{soft pPb}).
In the case of interaction with a nuclear target,
the Multiple Scattering Theory (Gribov-Glauber Theory)
is used, which allows to treat the interaction with the nuclear
target as the superposition of interactions
with different numbers of target nucleons (see a more detailed description
in~\cite{paper}, and references therein).

The average value of the number of target nucleons with whom the proton interacts, $\nu$,
has the well-known form:
\begin{equation}
\langle \nu \rangle = \frac{A\cdot\sigma^{pp}_{inel}}{\sigma^{pA}_{prod}} \;.
\end{equation}
We use the numerical values $\sigma^{pp}_{inel} \simeq 72$~mb
and $\sigma^{pPb}_{prod} \simeq 1900$~mb
at $\sqrt{s_{NN}}$=5~TeV, so that
\begin{equation}
\label{7.9}
\langle \nu \rangle_{p+Pb}\approx 7.9.
\end{equation}

In the calculation of the inclusive spectra of secondaries produced
in $pA$ collisions, the possibility of one or several Pomeron cuts in each
of the $\nu$ blobs of the proton-nucleon inelastic interactions should be considered.

The QGSM gives a reasonable description \cite{KTMS,Sh4}
of the inclusive
spectra of different secondaries produced in hadron-nucleus
collisions at energies $\sqrt{s_{NN}}$=14$-$30~GeV.

The situation drastically changes at RHIC energies, where, from a theoretical point of view,
the authors of ref.~\cite{CKTr} claimed
that the suppression factor in the inclusive density for $Pb$-$Pb$ collisions
when taken into account saturation effects was of about 2.
Later, this effect was experimentally confirmed
when comparing the theoretical inclusive densities
without saturation effects to the corresponding RHIC
experimental data for $Au$-$Au$ collisions~\cite{Phob,Phen}.

However, all estimations are model dependent.
In particular, the calculations of inclusive densities and multiplicities,
both in $pp$, and in heavy ion
collisions (with accounting
for inelastic nuclear screening),
can be fulfilled in the percolation theory~\cite{Dias}.

The percolation approach assumes two or several Pomerons
to overlap in the transverse space and to fuse in a single Pomeron.
Given a certain transverse radius, when the
number of Pomerons in the interaction region increases,
at least part of them may appear inside another Pomeron.
As a result, the internal partons (quarks and gluons) can split,
leading to the saturation of the final inclusive density. This effect
persists with the energy growth until all the Pomerons
will overlap~\cite{Dias}.

In order to account for the percolation effects in the QGSM,
it is technically more simple~\cite{MPS} to consider in
the central region the maximal number of
Pomerons $n_{max}$ emitted by one nucleon (see~\cite{MPS} for details).
By doing this, the QGSM calculations
of the spectra of different secondaries integrated
over $p_T$, as functions of initial energies, rapidity, and $x_F$,
become rather simple and very similar to those in the percolation approach.

In the following calculations, the value
$n_{max} = 21$ has been used at the LHC energy $\sqrt{s_{NN}}$=5~TeV. This value
can be regarded as the normalization of all the charged
secondaries multiplicities in the midrapidity region
to the ALICE data~\cite{ALICE}.

The predictive power of our calculation
applies for different sorts of secondaries in midrapidity region.
If the inelastic nuclear screening comes mainly from
the Pomeron interactions, as it was discussed above,
the screening effects would be the same for all the secondaries.

In the following calculations, one additional effect is also
taken into account, namely the transfer of the baryon charge to
large distances in rapidity space through the string junction
effect~\cite{ACKS,MRS}.
This transfer leads to an asymmetry in the production
of baryons and antibaryons in the central region
that is non-zero even at LHC energies (see~\cite{MRS} for the details of
the calculation of these effects).

\section{Rapidity spectra of different secondaries at LHC energies}

To compare the calculated effect of nuclear
screening with the experimental data, the adequate description
of the secondary production on nucleon, as well as on nuclear targets
is needed.

First, we have obtained the QGSM description of $\pi^\pm$,
$K^\pm$, $p$, and $\overline{p}$ productions in $pp$ collisions at LHC energies,
and then we have compared the results of our calculations with the experimental data by
the CMS Collaboration~\cite{CMS,CMSpp}
and by the ALICE Collaboration~\cite{ALICEpp09t,ALICEpp276t,ALICEpp7t}.
The experimental data by the ALICE Collaboration are
approximately 20$-$30\% lower than those published by the CMS Collaboration,
but in spite of this disagreement between ALICE and CMS data our QGSM result is qualitatively
compatible with both experimental samples (see~\cite{paper} for
the details of this analysis).

One has to note that the experimental point by the ALICE Collaboration~\cite{ALICE},
$dn_{ch}/d\eta = 16.81 \pm 0.71$
at $\sqrt{s_{NN}}$=5~TeV, has been used~\cite{soft pPb} to normalize the QGSM calculations
for the case of nuclear targets, the agreement of our calculations with this result being reached at
$n_{max} =21$, where the theoretical value is
of $dn_{ch}/d\eta=16.28$ (see ref.~\cite{soft pPb}).

The experimental data for $p$+Pb collisions by the CMS Collaboration on the
inclusive densities of different secondaries, $\pi^\pm$,
$K^\pm$, $p$, and $\overline{p}$~\cite{CMS} are presented in
Table~1, where they are compared with the results of our QGSM calculations.
The agreement for every secondary particles is good, what it
means that the experimental nuclear shadowing factor
is the same for different secondaries, as it is assumed in our
calculations.

Also in Table~1, we present the QGSM results for
the $pp$ collisions at the same energy.
The ratios of particle yields in $p$+Pb and $pp$ collisions
are equal to 3.6$-$3.7, i.e they are two times smaller than
the values of $\nu_{p+Pb}$ in Eq.~\ref{7.9}.
In the absence of inelastic nuclear screening, the ratio
$r =pPb/pp$ in the midrapidity region should be equal
to $\nu_{p+Pb}$~\cite{Sh3},
that is, to the average number of the inelastic collisions
of the incident proton in the target nucleus.
Thus, we can see that the inelastic nuclear screening factor
is little larger than 2, and it is practically the same
for all considered secondaries.

We have also calculated the hyperon and antihyperons production
in $pp$ and $p$+Pb collisions at the same energy $\sqrt{s_{NN}}$=5~TeV.
The ratios of the inclusive densities of all secondary
hyperons and antihyperons
produced on Pb and hydrogen targets are practically the same
as for secondary mesons production, with a $\sim 5\%$ accuracy (see~\cite{paper}),
what would indicate that the main contribution to the processes of hyperon
and meson production has a similar nature.

\begin{center}
\vskip 5pt
\begin{tabular}{|c||c||c|c|c|} \hline
particles &CMS Collaboration & \multicolumn{3}{c|}{QGSM} \\
\cline{3-5}
&  $dn/dy$, $|y|\leq 1$~\cite{CMS} &       $p$+Pb &  $pp$ & $r$\\
\hline
\hline
$\pi^+$  & $8.074 \pm 0.087$ & 8.103 & 2.190 & 3.70\\ \hline

$\pi^- $ & $7.971 \pm 0.079 $ & 7.923 & 2.147 & 3.69  \\ \hline

$K^+$ & $1.071 \pm 0.069 $ &1.006 &0.273 & 3.69 \\ \hline

$K^-$ & $0.984 \pm 0.047 $ &0.996 &0.271 & 3.66 \\ \hline

$p$ & $0.510 \pm 0.018 $ &0.545 &0.150 & 3.63 \\ \hline

$\bar p$ & $0.494 \pm 0.017 $ &0.536 &0.148 & 3.62  \\ \hline

\hline
\end{tabular}
\end{center}
Table~1: {\footnotesize Experimental data on $dn/dy$, $|y|\leq 1$ by the
CMS Collaboration~\cite{CMS} of charged pions, kaons,
$p$, and $\overline{p}$ production in central $p$+Pb collisions
at $\sqrt{s_{NN}}$=5~TeV, together with
the corresponding QGSM results. The parameter
$r$ is the ratio of the particle yields in $p$+Pb and $pp$
reactions. The results of the QGSM calculations for $pp$ collisions are also given.}

\section{Conclusion}

The inelastic nuclear screening corrections at LHC energies have proved to be really large,
and the QGSM approach for high energy inelastic pp, p-nucleus and nucleus-nucleus
collisions with multiparticle production,
provides a natural explanation of the independence of
the nuclear screening effects on the type of the produced particles
in the central region of inclusive spectrum,
as the nuclear screening effects are practically the same
for $\pi^\pm$, $K^\pm$, $p$, and $\overline{p}$ production.

If confirmed experimentally for high energy inelastic pp, p-nucleus and nucleus-nucleus
collisions with multiparticle production, this fact would indicate that the interaction
of secondaries in the final state would be negligibly small.
\vskip 0.4cm

{\bf Acknowledgements}

The authors thank B.Z.~Kopeliovich for his valuable comments.

C.M. wants to congratulate Christophe Royon and all the organizers
for the nice environement for talks and discussion
at the conference, and to thank Nicola Minafra for his help during the whole stay in
Bisceglie.

This work has been supported by Russian RSCF grant No. 14-22-00281,
by the State Committee of Science of the Republic of Armenia, Grant-15T-1C223,
by Ministerio de Ciencia e Innovaci\'on of Spain under project
FPA2014-58293-C2-1-P and FEDER, and the Spanish Consolider-Ingenio 2010 Programme CPAN (CSD2007-00042),
and by Xunta de Galicia, Spain (2011/PC043).

\end {document}